\documentclass[Physsubmission, Phys]{SciPost}

\usepackage{url}
\usepackage{hyperref}
\definecolor{nicered}{rgb}{0.5,0.,0.}
\definecolor{nicegreen}{rgb}{0.,0.5,0.}
\definecolor{niceblue}{rgb}{0.,0.,0.5}
\hypersetup{colorlinks,citecolor=nicegreen,linkcolor=nicered,urlcolor=niceblue}

\usepackage[bitstream-charter]{mathdesign}
\DeclareSymbolFont{usualmathcal}{OMS}{cmsy}{m}{n}
\DeclareSymbolFontAlphabet{\mathcal}{usualmathcal}

\begin{document}

\begin{center}{\Large \textbf{
Nucleon and pion PDFs: large-x asymptotics meets functional mimicry
}}\end{center}

\begin{center}
A. Courtoy\textsuperscript{1} and
P. Nadolsky\textsuperscript{2$\star$}
\end{center}

\begin{center}
{\bf 1} Instituto de F\'isica, Universidad Nacional Aut\'onoma de M\'exico\\
Apartado Postal 20-364, 01000 Ciudad de M\'exico, Mexico
\\
{\bf 2} Department of Physics, Southern Methodist University, Dallas, TX 75275-0181, U.S.A.
\\

* aurore@fisica.unam.mx
\end{center}

\begin{center}
\today
\end{center}


\definecolor{palegray}{gray}{0.95}
\begin{center}
\colorbox{palegray}{
  \begin{tabular}{rr}
  \begin{minipage}{0.1\textwidth}
    \includegraphics[width=22mm]{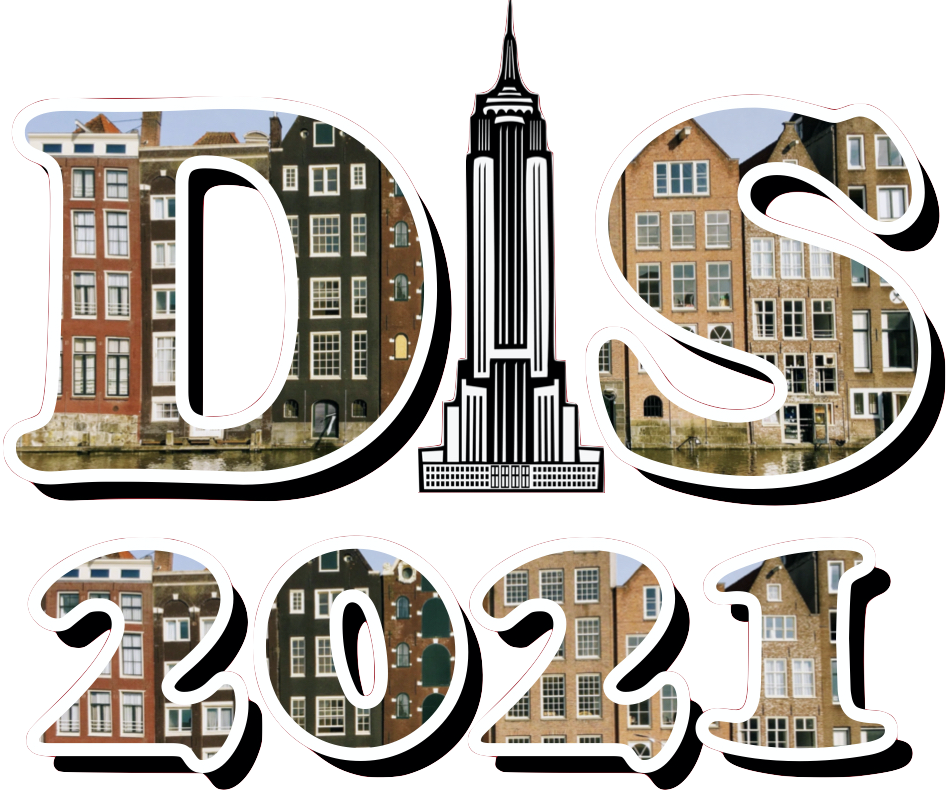}
  \end{minipage}
  &
  \begin{minipage}{0.75\textwidth}
    \begin{center}
    {\it Proceedings for the XXVIII International Workshop\\ on Deep-Inelastic Scattering and
Related Subjects,}\\
    {\it Stony Brook University, New York, USA, 12-16 April 2021} \\
    \doi{10.21468/SciPostPhysProc.?}\\
    \end{center}
  \end{minipage}
\end{tabular}
}
\end{center}

\section*{Abstract}
{\bf
We review strategies to unveil the primordial large-$x$ structure of the nucleons as well as the pion from hard-scattering experiments. Ideas are presented for learning about the $x\to 1$ limit of nonperturbative QCD dynamics at energy scales of order 1 GeV from collider experiments at much higher scales. The behavior of parton distributions at $x\to 1$ predicted by the quark counting rules and other  low-energy theoretical approaches is contrasted with phenomenological PDFs. Polynomial mimicry of PDF parametrizations is one of many factors that influence the apparent power of the $(1-x)$ falloff. We discuss implications of the mimicry for the large-$x$ falloff of the pion PDFs.
}


\section{Introduction}
\label{sec:intro}

Recent analyses of the pion structure on the lattice and in nonperturbative QCD approaches have drawn attention to the behavior of parton distribution functions (PDFs) at large momentum fractions $x$. Early QCD models~\cite{Ezawa:1974wm,Farrar:1975yb,Berger:1979du,Soper:1976jc}
predict a fall-off of structure functions of $(1-x)^{2 n_s-1+2|\lambda_q-\lambda_A|}$,  where $n_s$ is the number of quark spectators, and $\lambda_{q,A}$ the helicity of the active quark and parent hadron, respectively. 

On the side of nucleon phenomenology,
the asymptotic behavior of PDFs in the $x\to 1$ limit is best known for unpolarized protons; the $(1-x)^3$ fall-off of their valence PDFs has long been consistent with PDF parametrizations found in global QCD analyses. Recent detailed phenomenological studies~\cite{Ball:2016spl,Courtoy:2020fex} concluded that the scarcity of the observational data at large $x$ results in a large uncertainty about the proton's falloff dependence. From the point of view of low-energy models, the dynamics inside the proton is predominantly simulated by confining models, few of which have considered the shape of the PDF at large $x$.

On the other hand, the pion structure is understood to be dominated by chiral symmetry breaking, which, together with pion's two-body nature, allowed
for first-principle and field-theoretical predictions, such as in the Dyson-Schwinger formalism ({\it e.g.}~\cite{Ding:2019lwe,Bednar:2018mtf}) and lattice QCD ({\it e.g.}~\cite{Zhang:2018nsy,Gao:2020ito,Alexandrou:2021mmi}).
These suggest that  manifestations of low-energy QCD dynamics may be seen in the shape of pion PDFs at $x>0.5$. Phenomenological parametrizations of pion PDFs, {\it e.g.} \cite{Barry:2018ort,Novikov:2020snp}, serve as a mediator between theoretical approaches and observations. Yet, evidence for the predicted large-$x$ falloff $(1-x)^2$ of the pion PDFs is ambiguous. 
The newest results further stimulated discussions about the interpretation of QCD-based arguments in view of the behavior of  PDFs at large $x$~\cite{Gao:2020ito,Broniowski:2020had,Courtoy:2020fex}.

In Ref.~\cite{Courtoy:2020fex}, we explored the applicability of the quark counting rules in modern global QCD analyses. Conditions under which the counting rules hold deviate from those in phenomenologically accessible observables for various reasons, including the difference between the DIS structure functions -- for which the counting rules were demonstrated -- and $\overline{\rm MS}$ PDFs, the presence of threshold resummation at large $x$, or a {\it non-minimally} perturbed parent hadron. Control of such factors is crucial for bridging the  nonperturbative interpretation to the PDF models based on scattering data.
To these physical considerations must be added a mathematical one: is it possible to deduce the exact power law for a PDF from the PDF's polynomial form fitted to the empirical data?

\section{Polynomial mimicry}

Whether it is based on experimental data or Mellin moment reconstructions, the determination of PDFs beyond analytical results 
relies on polynomial parametrizations more often than not. In such cases, discrete data can be compatible with more than just one functional form, {\it i.e.} there is no unique form that proves that data demand a $(1-x)^{A_2}$ fall-off, where $A_2$ would be larger than $3$ for a proton and $2$ for a pion.
\begin{figure}[h]
\centering
\includegraphics[width=0.325\textwidth]{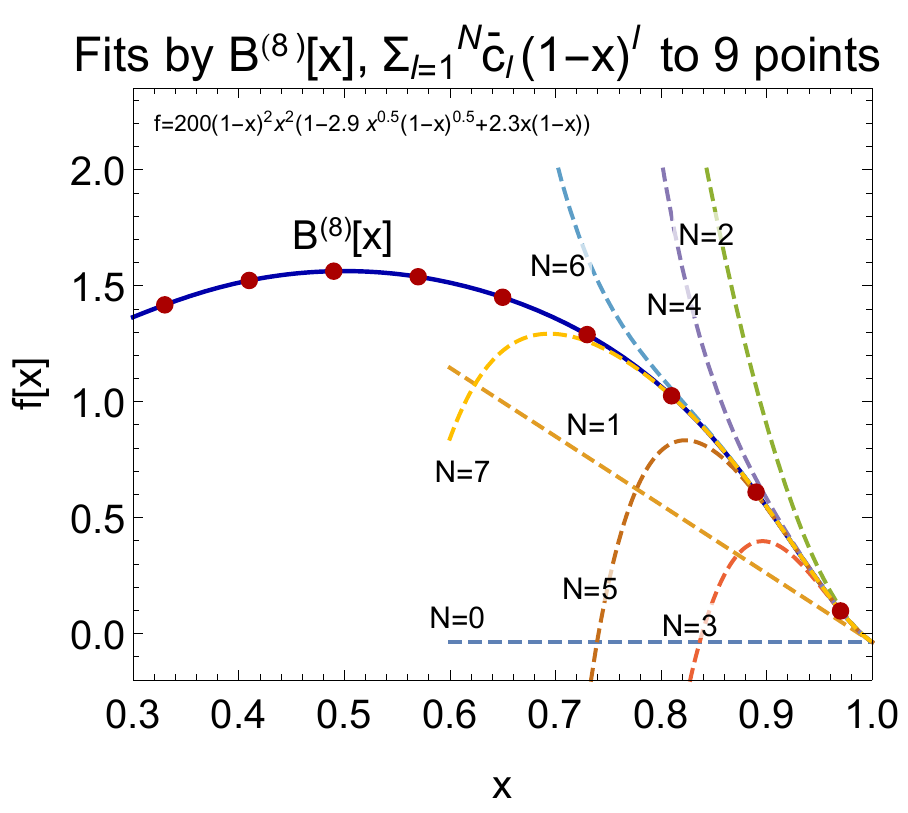}
\includegraphics[width=0.325\textwidth]{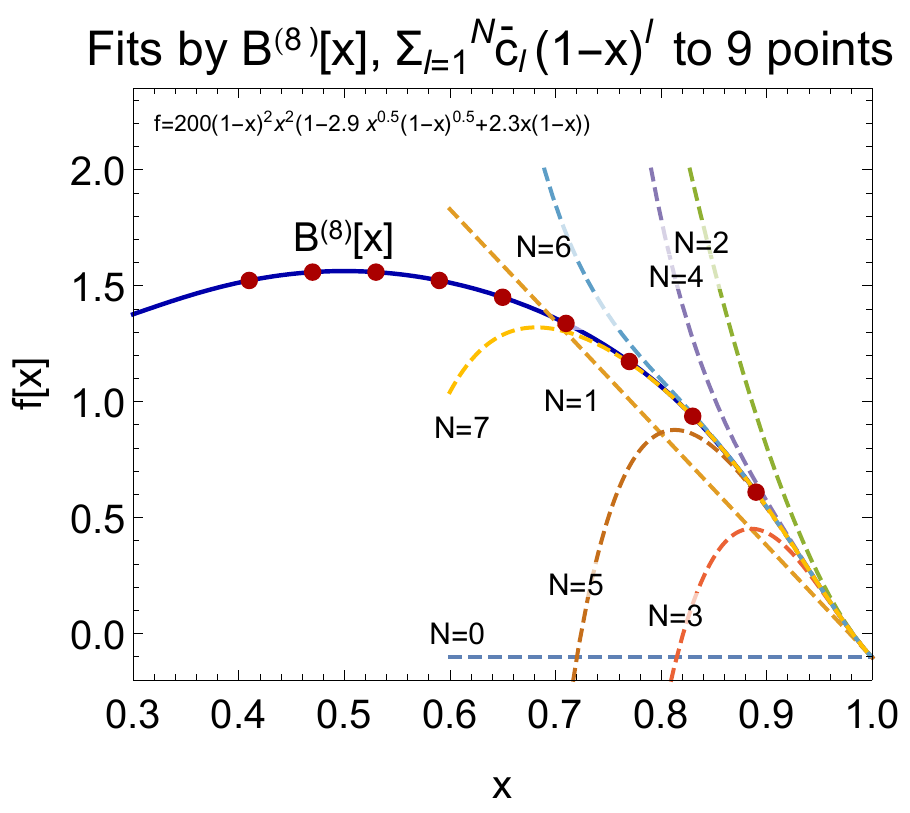}
\includegraphics[width=0.325\textwidth]{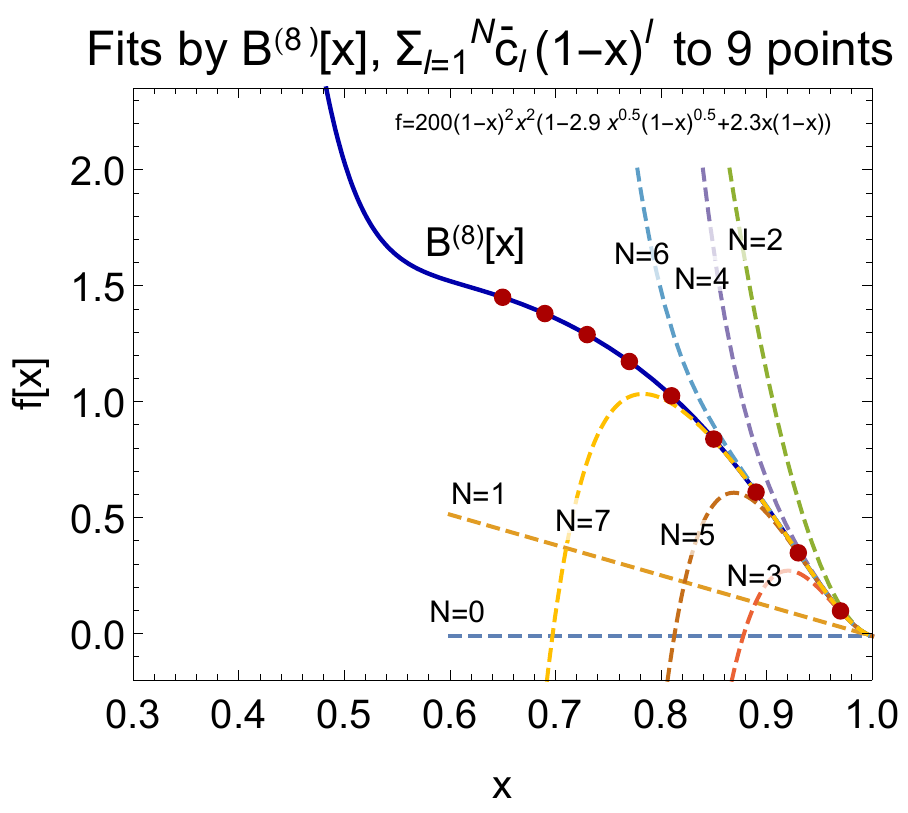}
\caption{B\'ezier polynomial for the reconstruction of the function $f(x)=200x^2(1-x)^2(1-2.9\sqrt{x(1-x)}+2.3x(1-x))$ from 9 control points spaced differently in $x$ on each subplot. The dashed curves correspond to the monomial expansions on the right-hand side in Eq.~(\ref{eq:Bnonemx}) truncated at $n=N$.}
\label{fig:monomial}
\end{figure}

To demonstrate the latter statement, in Ref.~\cite{Courtoy:2020fex}
we have explored how the  falloff exponent of a model pion PDF depends on the assumed polynomial form that interpolates a set of discrete data points at which the pion PDF is sampled. We used exact interpolations based on B\'ezier curves, each rendering a unique polynomial solution for the chosen set of $(n+1)$ sampled points. A strategy to determine
the falloff of PDFs as $x\to 1$ relies on the monomial expansion of the reconstructed B\'ezier curve
\begin{eqnarray}
 {\cal B}^{(n)}(x) = \sum_{l=0}^n \bar{c}_l\  (1-x)^l.
  \label{eq:Bnonemx}
\end{eqnarray}
For realistic functional forms, we observe a mixing of expansion coefficients, $\bar{c}_l$. In Fig.~\ref{fig:monomial}, it can be appreciated that the lowest coefficients of the monomial expansion, notably the linear $N=1$ term, are spurious and depend on the range and spacing of the sampled data.  It means that, because of the mimicry, polynomial forms can and will contain solutions that might not necessarily agree with the true power law, as the lowest powers of the expansion cannot be meaningfully reconstructed.

For such pion PDFs, data at $x\gtrsim 0.9$ are necessary to minimize that mixing. In that sense, it is worth mentioning the conceptual difference between the theoretical $x\to 1$ limit and the phenomenological definition at {\it large $x$}, see Sec.~\ref{sec:A2effColliders}. Polynomial mimicry should translate into an increase of the uncertainty based on the multiple choices for the functional forms in the region $x\to 1$.

As a consequence, due to mimicry, it becomes difficult to disentangle the manifestations of nonperturbative dynamics from the functional-form uncertainties and other effects on the shape of the PDFs found from realistic data.

\section{Testing the large-$x$ PDF falloff at colliders \label{sec:A2effColliders}}
\begin{figure}[h]
\centering
\includegraphics[width=0.48\textwidth]{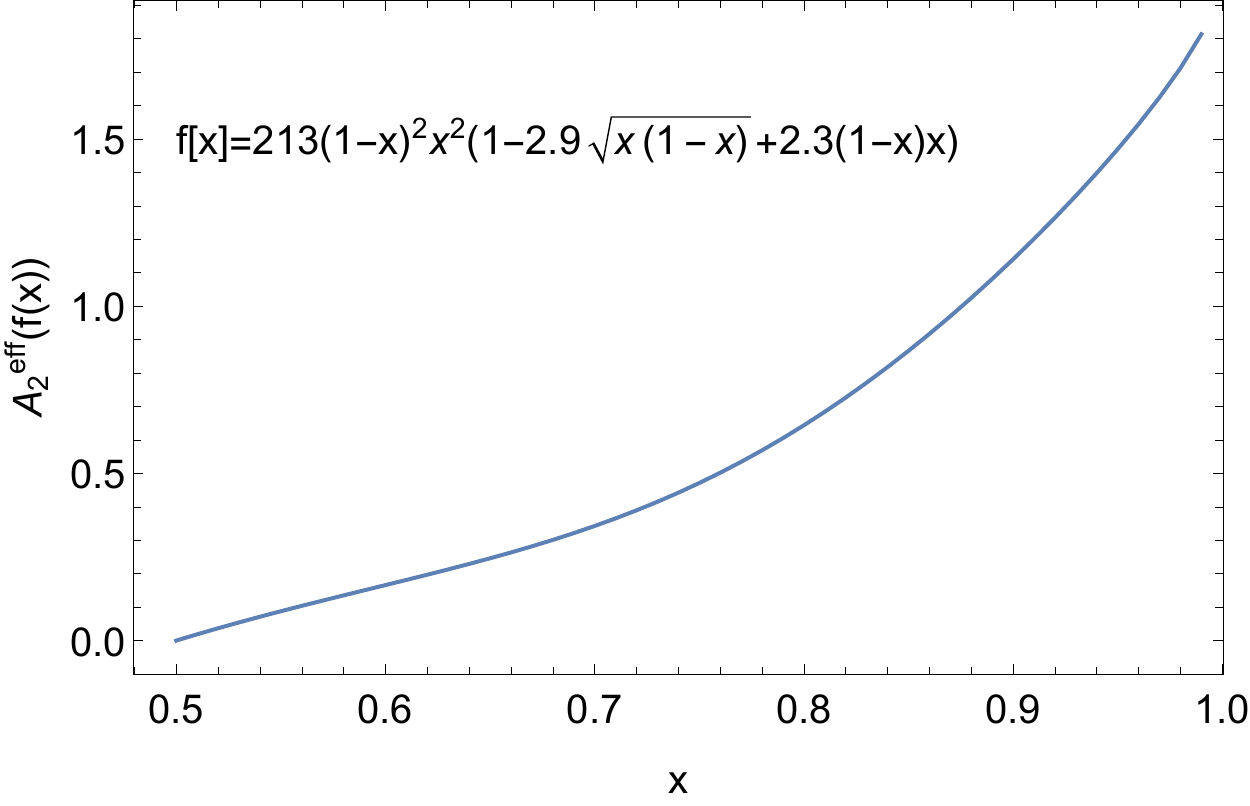}\quad
\includegraphics[width=0.48\textwidth]{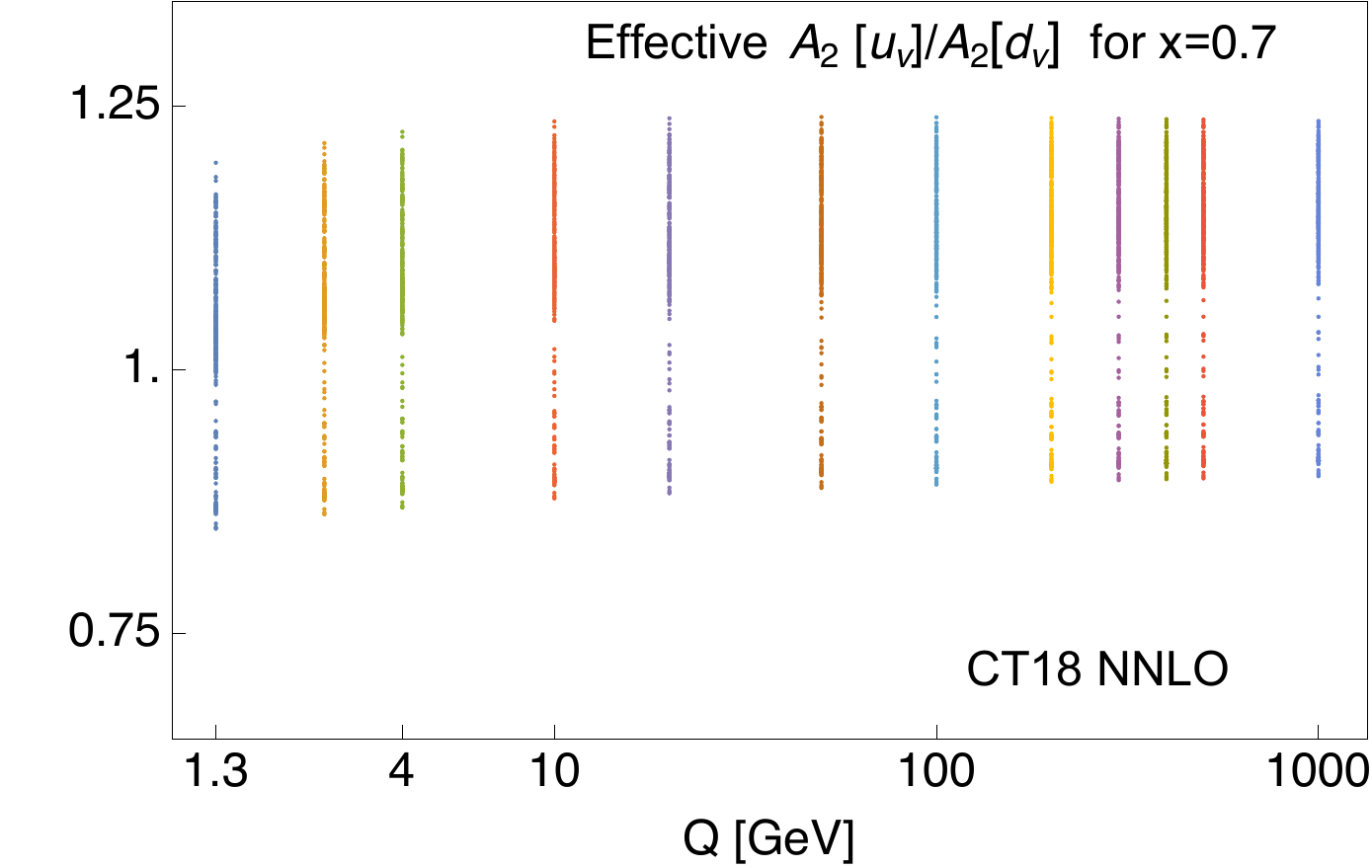}
\caption{Left: The effective $A_2^{\rm eff}(x)$ for a toy pion parametrization $f(x)\propto (1-x)^{A_2}$ with $A_2=2$. Right: Ratio $A_2^{\rm eff}[u_V]/A_2^{\rm eff}[d_V]$ for CT18NNLO valence PDFs at $x=0.7$ as a function of the scale $Q$.}
\label{fig:uvOvdV}
\end{figure}

A more reliable strategy to access the $(1-x)$-exponent is found by observing that, for $x\to 1$, most nonperturbative forms behave as 
\begin{equation}
{\cal F}(x, Q^2)=(1-x)^{A_{2}}\times \Phi(1-x)
\;,
\end{equation}
where $\Phi(1-x)$ is a relatively slow function. It is indeed the form used by several global analyses, in particular CT18NNLO~\cite{Hou:2019efy}, and it will be used here below. We therefore advocate to study the effective behavior of the large-$x$ logarithmic derivative
\begin{equation}
A_2^{\rm eff}\left[{\cal F}(x,Q^2)\right]\equiv \frac{\partial \ln\left({\cal F}(x, Q^2)\right)}{\partial \ln\left(1-x\right)}=A_2\left[{\cal F}\right] +\mbox{correction}
\;,
\end{equation}
from which we expect to recover $A_2^{\rm  eff}=A_2$ for $x\sim 1$. 
The difference between the effective and true exponents depends on $x$ and can still be large for PDF parametrizations with concurring dynamical effects, see an example for the pion PDF in the left Fig.~\ref{fig:uvOvdV}.

The effective exponents for the CT18  ensemble of NNLO proton PDFs \cite{Hou:2019efy} have been studied in Ref.~\cite{Courtoy:2020fex}. The DIS structure function $F_2(x,Q^2)$ was found to be compatible with the quark counting rule expectation of $A_2^{\rm eff}(F_2)=3$ as $x\to 1$, within error bands. As mentioned above, multiple factors contribute to the PDF uncertainty~\cite{Kovarik:2019xvh}, and the error bands are large in the regions with scarce data, such as $x\to 1$. 
\begin{figure}[h]
\centering
\includegraphics[width=0.85\textwidth]{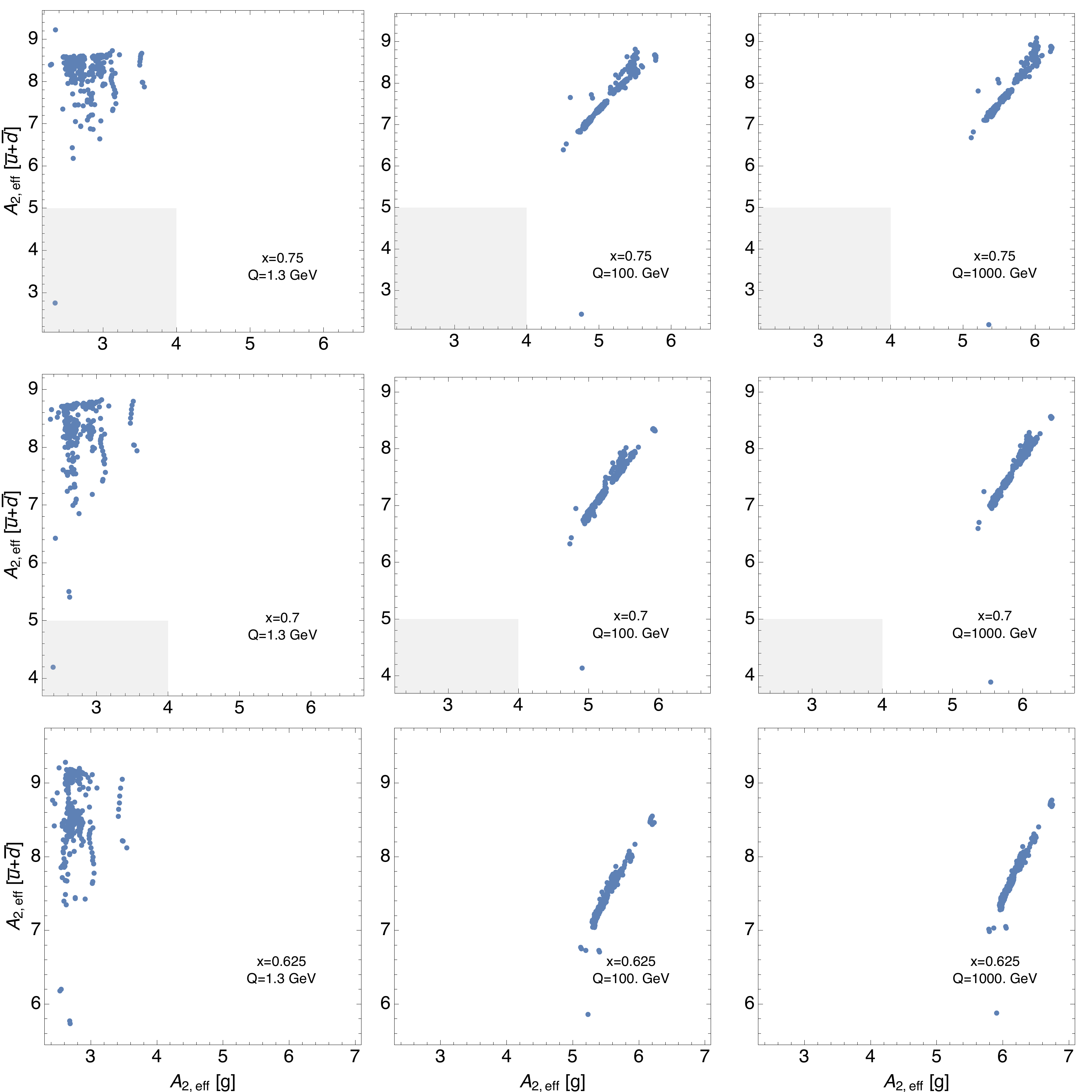}
\caption{$A_2^{\rm eff}$ for the gluon against  $\bar{u}+\bar{d}$, for various $x$ values. The $Q$ values, from left to right, $1.3, \, 100$, and $1000$ GeV. The shaded area corresponds to regions unfavored by quark counting rules.}
\label{fig:glvssea}
\end{figure}

The effective exponents of the PDFs evolve with $Q^2$ in accordance with DGLAP evolution, which is known very accurately to NNLO in $\alpha_s$.  Knowledge of QCD evolution opens interesting opportunities for testing $A_2^{\rm eff}$ at nonperturbative $Q\sim 1$ GeV by using large-$x$ constraints at very high $Q$ from collider experiments, such as the ZEUS measurements \cite{ZEUS:2020ddd} of DIS at $Q\approx 100$ GeV.

Of particular interest for testing nonperturbative dynamics is the $d/u$ ratio (see contribution from C. Keppel in these proceedings, as well as, {\it e.g.}, Ref.~\cite{Accardi:2021ysh}). 
As both the up and down valence PDFs evolve according to the same non-singlet DGLAP evolution, the $Q^2$ dependence of their effective exponents cancels in a wide $x$ range. Figure~\ref{fig:uvOvdV} shows the ratio $A_{2,u_{\rm V}}^{\rm eff}(x)/A_{2,d_{\rm V}}^{\rm eff}(x)$ for 363 trial parametrizations of CT18 NNLO PDFs versus $Q$. The CT18 parametrizations assume the ratio $A_{2,u_{\rm V}}/A_{2,d_{\rm V}}$ of the true exponents to be equal to $1$ to ensure that the $d/u$ remains finite. The effective exponents at $x < 1$  deviate from the true ones. The ratio is nearly invariant under DGLAP evolution, so it can be determined at ZEUS or LHC and provide accurate insights about the flavor composition in the valence sector at $Q\sim 1$ GeV. 

Figure~\ref{fig:glvssea} shows that the rate of the $Q^2$ evolution for the poorly known gluon $A_{2,g}^{\rm eff}$ depends only weakly on the PDF parametrization form. Here we show the clouds of $A_{2,\bar u+\bar d}^{\rm eff}$ vs. $A_{2,g}^{\rm eff}$ for 363 trial CT18 NNLO parametrizations at the specified $x$ and $Q$, in the same fashion as in Ref.~\cite{Courtoy:2020fex} but for $Q$ extending to 1000 GeV. Constraints on $A_{2,g}^{\rm eff}$ in jet production at the LHC could be evolved backward to low $Q$ to provide insights about the nonperturbative dynamics.

\section{Conclusion}
In these proceedings, we have summarized our strategies for exploring the large-$x$ behavior of PDFs. Due to the polynomial mimicry, it is not possible to uniquely determine the exact $(1-x)$-exponent given the present data. A similar conclusion has been reached in some lattice studies~\cite{Broniowski:2020had, Gao:2020ito}. Our second strategy consists in studying the effective exponent, expressed as a logarithmic derivative of the structure function or PDF. The effective falloff can be determined from a combination of low-$Q$ and high-$Q$ data and confronted with its computation in a nonperturbative QCD model. This approach bridges the first-principle predictions with realistic observations.

\section*{Acknowledgements}
The authors thank A.~Caldwell, M.~Crayencour, T.~J.~Hobbs for discussions as well as Xiang Gao for pointing out the study of the $(1-x)$-exponent in Ref.~\cite{Gao:2020ito}.


\paragraph{Funding information}
AC is supported by UNAM Grant No. DGAPA-PAPIIT IA101720 and CONACyT Ciencia de Frontera 2019 No.~51244 (FORDECYT-PRONACES). PN is partially supported by the U.S.~Department of Energy under Grant No.~DE-SC0010129.



\bibliography{ct18bibtex.bib}

\begin{thebibliography}{10}
\providecommand{\url}[1]{\texttt{#1}}
\providecommand{\urlprefix}{URL }
\expandafter\ifx\csname urlstyle\endcsname\relax
  \providecommand{\doi}[1]{doi:\discretionary{}{}{}#1}\else
  \providecommand{\doi}{doi:\discretionary{}{}{}\begingroup
  \urlstyle{rm}\Url}\fi
\providecommand{\eprint}[2][]{\url{#2}}

\bibitem{Ezawa:1974wm}
Z.~Ezawa,
\newblock \emph{{Wide-Angle Scattering in Softened Field Theory}},
\newblock Nuovo Cim. A \textbf{23}, 271 (1974),
\newblock \doi{10.1007/BF02739483}.

\bibitem{Farrar:1975yb}
G.~R. Farrar and D.~R. Jackson,
\newblock \emph{{Pion and Nucleon Structure Functions Near x=1}},
\newblock Phys. Rev. Lett. \textbf{35}, 1416 (1975),
\newblock \doi{10.1103/PhysRevLett.35.1416}.

\bibitem{Berger:1979du}
E.~L. Berger and S.~J. Brodsky,
\newblock \emph{{Quark Structure Functions of Mesons and the Drell-Yan
  Process}},
\newblock Phys. Rev. Lett. \textbf{42}, 940 (1979),
\newblock \doi{10.1103/PhysRevLett.42.940}.

\bibitem{Soper:1976jc}
D.~E. Soper,
\newblock \emph{{The Parton Model and the Bethe-Salpeter Wave Function}},
\newblock Phys. Rev. D \textbf{15}, 1141 (1977),
\newblock \doi{10.1103/PhysRevD.15.1141}.

\bibitem{Ball:2016spl}
R.~D. Ball, E.~R. Nocera and J.~Rojo,
\newblock \emph{{The asymptotic behaviour of parton distributions at small and
  large $x$}},
\newblock Eur. Phys. J. C \textbf{76}(7), 383 (2016),
\newblock \doi{10.1140/epjc/s10052-016-4240-4},
\newblock \eprint{1604.00024}.

\bibitem{Courtoy:2020fex}
A.~Courtoy and P.~M. Nadolsky,
\newblock \emph{{Testing momentum dependence of the nonperturbative hadron
  structure in a global QCD analysis}},
\newblock Phys. Rev. D \textbf{103}(5), 054029 (2021),
\newblock \doi{10.1103/PhysRevD.103.054029},
\newblock \eprint{2011.10078}.

\bibitem{Ding:2019lwe}
M.~Ding, K.~Raya, D.~Binosi, L.~Chang, C.~D. Roberts and S.~M. Schmidt,
\newblock \emph{{Symmetry, symmetry breaking, and pion parton distributions}},
\newblock Phys. Rev. D \textbf{101}(5), 054014 (2020),
\newblock \doi{10.1103/PhysRevD.101.054014},
\newblock \eprint{1905.05208}.

\bibitem{Bednar:2018mtf}
K.~D. Bednar, I.~C. Clo\"et and P.~C. Tandy,
\newblock \emph{{Distinguishing Quarks and Gluons in Pion and Kaon Parton
  Distribution Functions}},
\newblock Phys. Rev. Lett. \textbf{124}(4), 042002 (2020),
\newblock \doi{10.1103/PhysRevLett.124.042002},
\newblock \eprint{1811.12310}.

\bibitem{Zhang:2018nsy}
J.-H. Zhang, J.-W. Chen, L.~Jin, H.-W. Lin, A.~Sch\"afer and Y.~Zhao,
\newblock \emph{{First direct lattice-QCD calculation of the $x$-dependence of
  the pion parton distribution function}},
\newblock Phys. Rev. D \textbf{100}(3), 034505 (2019),
\newblock \doi{10.1103/PhysRevD.100.034505},
\newblock \eprint{1804.01483}.

\bibitem{Gao:2020ito}
X.~Gao, L.~Jin, C.~Kallidonis, N.~Karthik, S.~Mukherjee, P.~Petreczky,
  C.~Shugert, S.~Syritsyn and Y.~Zhao,
\newblock \emph{{Valence parton distribution of the pion from lattice QCD:
  Approaching the continuum limit}},
\newblock Phys. Rev. D \textbf{102}(9), 094513 (2020),
\newblock \doi{10.1103/PhysRevD.102.094513},
\newblock \eprint{2007.06590}.

\bibitem{Alexandrou:2021mmi}
C.~Alexandrou, S.~Bacchio, I.~Clo\"et, M.~Constantinou, K.~Hadjiyiannakou,
  G.~Koutsou and C.~Lauer,
\newblock \emph{{The pion and kaon $\langle x^3 \rangle$ from lattice QCD and
  PDF reconstruction from Mellin moments}}  (2021),
\newblock \eprint{2104.02247}.

\bibitem{Barry:2018ort}
P.~Barry, N.~Sato, W.~Melnitchouk and C.-R. Ji,
\newblock \emph{{First Monte Carlo Global QCD Analysis of Pion Parton
  Distributions}},
\newblock Phys. Rev. Lett. \textbf{121}(15), 152001 (2018),
\newblock \doi{10.1103/PhysRevLett.121.152001},
\newblock \eprint{1804.01965}.

\bibitem{Novikov:2020snp}
I.~Novikov \emph{et~al.},
\newblock \emph{{Parton Distribution Functions of the Charged Pion Within The
  xFitter Framework}},
\newblock Phys. Rev. D \textbf{102}(1), 014040 (2020),
\newblock \doi{10.1103/PhysRevD.102.014040},
\newblock \eprint{2002.02902}.

\bibitem{Broniowski:2020had}
W.~Broniowski and E.~Ruiz~Arriola,
\newblock \emph{{Vector-axial vector lattice cross section and valence parton
  distribution in the pion from a chiral quark model}},
\newblock Phys. Lett. \textbf{B}, 135803 (2020),
\newblock \doi{10.1016/j.physletb.2020.135803},
\newblock \eprint{2006.03832}.

\bibitem{Hou:2019efy}
T.-J. Hou \emph{et~al.},
\newblock \emph{{New CTEQ global analysis of quantum chromodynamics with
  high-precision data from the LHC}},
\newblock Phys. Rev. D \textbf{103}(1), 014013 (2021),
\newblock \doi{10.1103/PhysRevD.103.014013},
\newblock \eprint{1912.10053}.

\bibitem{Kovarik:2019xvh}
K.~Kova\v{r}\'\i{}k, P.~M. Nadolsky and D.~E. Soper,
\newblock \emph{{Hadronic structure in high-energy collisions}},
\newblock Rev. Mod. Phys. \textbf{92}(4), 045003 (2020),
\newblock \doi{10.1103/RevModPhys.92.045003},
\newblock \eprint{1905.06957}.

\bibitem{ZEUS:2020ddd}
I.~Abt \emph{et~al.},
\newblock \emph{{Study of proton parton distribution functions at high $x$
  using ZEUS data}},
\newblock Phys. Rev. D \textbf{101}(11), 112009 (2020),
\newblock \doi{10.1103/PhysRevD.101.112009},
\newblock \eprint{2003.08742}.

\bibitem{Accardi:2021ysh}
A.~Accardi, T.~J. Hobbs, X.~Jing and P.~M. Nadolsky,
\newblock \emph{{Deuterium scattering experiments in CTEQ global QCD analyses:
  a comparative investigation}},
\newblock Eur. Phys. J. C \textbf{81}(7), 603 (2021),
\newblock \doi{10.1140/epjc/s10052-021-09318-y},
\newblock \eprint{2102.01107}.

\end{thebibliography}

\nolinenumbers

\end{document}